# Unveiling Parkinson's Disease-like Changes Triggered by Spaceflight

Nilufar Ali[*#], Afshin Beheshti[$], Greg Hampikian[#]

[#]Department of Biological Science, Boise State University, Boise, ID, 83725, USA

[$]Blue Marble Space Institute of Science, Seattle, WA, 98104 USA

[$]Stanley Center for Psychiatric Research, Broad Institute of MIT and Harvard, Cambridge, MA, 02142, USA

[*]Corresponding author nilufarali@boisestate.edu

**Abstract:**

A meta study of spaceflight data from both mouse and human flights reveals a striking overlap with Parkinson's disease (PD). Parallels include: changes in gait, loss of dopamine, sustained changes in the basal ganglia, loss of tyrosine hydroxylase in the substantia nigra, and systemic mitochondrial dysfunction. We identified specific Parkinson's genes differentially expressed post-spaceflight. These evidences indicate that spaceflight stressor-induced changes in the brain may become permanent during deep space exploration, posing a risk of PD in astronauts.

**Introduction:**

Parkinson's disease (PD) is the most common neurodegenerative movement disorder. It is caused by atrophy of the mid brain Nigro-striatal projections, resulting in symptoms of resting tremor, rigidity, bradykinesia and postural instability [1]. Genetic factors account for <10% of the disease burden with significant environmental contribution [2]. However, it is not known how the exposome (lifestyle, diet, exterior environment etc.) influences PD pathogenesis [3].

Spaceflight-induced systemic changes are similar to PD, raising concerns about astronaut health on deep space missions. This parallel may be due to the exposome of spaceflight including ionizing radiation (from galactic cosmic rays and solar particle events), microgravity, hypoxia, hypothermia, hypercapnia, confinement and associated physiological/psychological responses.

**Anatomical similarities between SF and PD**
In the early stages of PD, perivascular spaces (PVS), a commonly observed anatomical structure around the basal ganglia and midbrain [4], are affected and enlarged, presenting PVS burden [4]. These same regions are thought to be most sensitive to spaceflight [5, 6], and are associated with PVS burden post long-duration (but not short-duration) spaceflight [6]. Spaceflight induced changes in brain structures and circuitry linked to PD have been reported even after 7 months post-flight [7, 8]. These changes in brain architecture include a sustained grey matter increase in the Basal Ganglia (BG) [6, 8], which is also seen in PD cases [9]. Structural changes in other brain regions such as the cortex are reportedly reversed 6-7 months post spaceflight [10, 11].

**Change in gait and postural instability post spaceflight**
Gait characteristics can serve as indicators of vestibular, sensory and neuromotor dysfunction, as well as musculoskeletal atrophy resulting from diverse causes (including PD). Post-flight changes in astronaut gait have been extensively documented [12] and reflect deficits across multiple systems. Data from the Rodent Research (RR)-9 mission (35 days in space), revealed detailed alterations in mouse gait resembling those observed in movement disorders such as PD and Huntington's disease (HD) [12]. However, there was no follow up post-flight study for these animals as they were sacrificed within a day of return.

**Radiation exposure and PD**
Radiation exposure has been linked to PD [13, 14]. A meta-analysis was conducted from six cohorts in the Million Person Study (MPS) consisting of 517,608 American workers exposed to low-dose radiation, with maximum mean dose to the brain ranging from 0.76 to 2.7 Gy over a lifetime [14]. Five of the six cohorts had statistically significant positive associations with PD, based on 1573 deaths due to the disease [14]. This study indicates to us the need to assess the neurological implications of prolonged space travel and the benefits of specific radiation-protection measures. It should be noted that so far, the median time spent in space within Low-earth orbit by an astronaut is about a month, which leads to an accumulated radiation dosage of 7.52 mGy. NASA has set the radiation limit for astronauts on deep space missions (such as to the moon and Mars) at 1.5 mGy/day [15]. For the projected 900-day mission to Mars, the maximum dosage would be about 1 Gy, which is within the range of the MPS study that showed an increased risk of PD.

**Mitochondrial dysfunction**
Extensive research on astronauts and animals has revealed that exposure to spaceflight stressors can lead to accelerated aging, central nervous system impairments, and systemic mitochondrial dysfunction (MD) [16]. A landmark multi-omics study involving data from 59 astronauts and space-flown mice revealed that mitochondrial stress is a key hub for driving systemic changes contributing to health risks post spaceflight [17]. Data from the NASA Twin Study on long term space missions demonstrated increased mitochondrial stress and higher levels of mitochondria in the blood, both signs of MD [17, 18].

We reanalyzed the Gene Set Enrichment Analysis (GSEA) data from NASA's GeneLab comparing space-flown human cells to ground controls [17], looking for PD-like molecular changes. We found 137 PD-associated mitochondrial genes that are differentially expressed **(Fig 1 A-C)**. Many of these genes (e.g. PRKN, PINK1, UCHL1, LRRK2, & TH) are associated with MD, which is a key hallmark of genetic and idiopathic PD [19, 20] **(Fig 1 B)**.

Additionally, comparing blood transcriptomics data from PD patients [21] with that of the Inspiration 4 (I4) mission astronaut blood PBMC transcriptomics data [22, 23] reveal a striking match of downregulation in genes and pathways that are also significantly downregulated in PD **(Fig 1 D-F)**. Similar to PD, OXPHOS and mitochondrial ribosome genes were found to be downregulated even after 82 days post return (R+82) from space. Although the I4 astronauts had a short stay of 3 days at space, they flew farther than the ISS or Hubble, at a distance of 590 kms and hence they got and increased radiation exposure, causing long lasting MD. Interestingly, familial PD genes such as PRKN, PINK1, UCHL1, LRRK2, DJ1, VPS35, SNCA and GBA were also altered immediately post return (R+1) with sustained alteration even after 82 days post recovery (R+82) when compared to Preflight levels (**Fig 1 G**).

MD in PD results in increased mitochondrial DNA damage and ROS; dysregulation of bioenergetic capacity, quality control, fusion-fission dynamics, and mitochondrial transport. These aberrations lead to the activation of cell death pathways [19, 20, 24-26]. Systemic MD is observed in PD patients' Substantia Nigra pars compacta (SNpc); as well as in muscle, lymphocytes, and platelets [20, 24]. This systemic MD mirrors observations seen post spaceflight [17].

Systemic failure in both proteostasis and ribostasis have been inferred post-spaceflight by alterations in ribosome assembly, mitochondrial function, and cytosolic translation pathways [17]. Emerging evidence suggests a link between MD and proteostasis failure involving the PINK1/Parkin pathways [27]. Age-related neurodegenerative diseases including PD also manifest proteostasis and ribostasis failure [28].

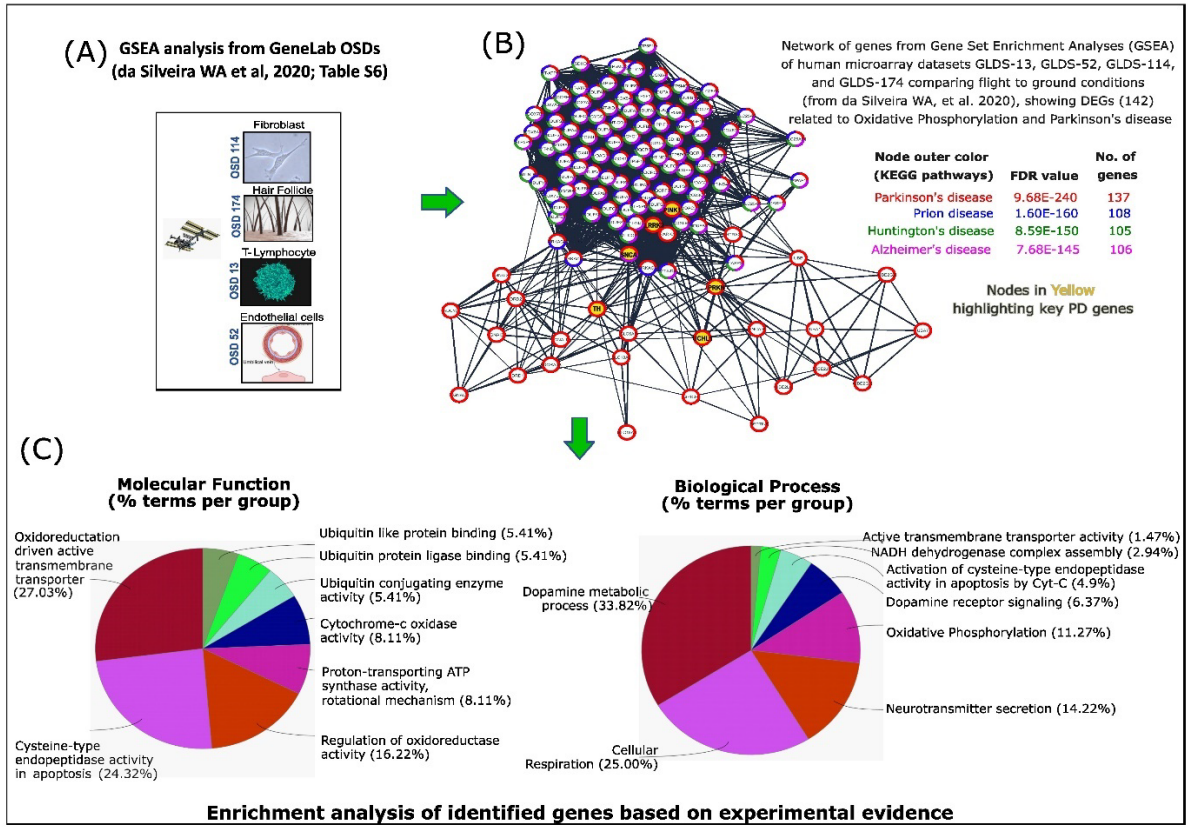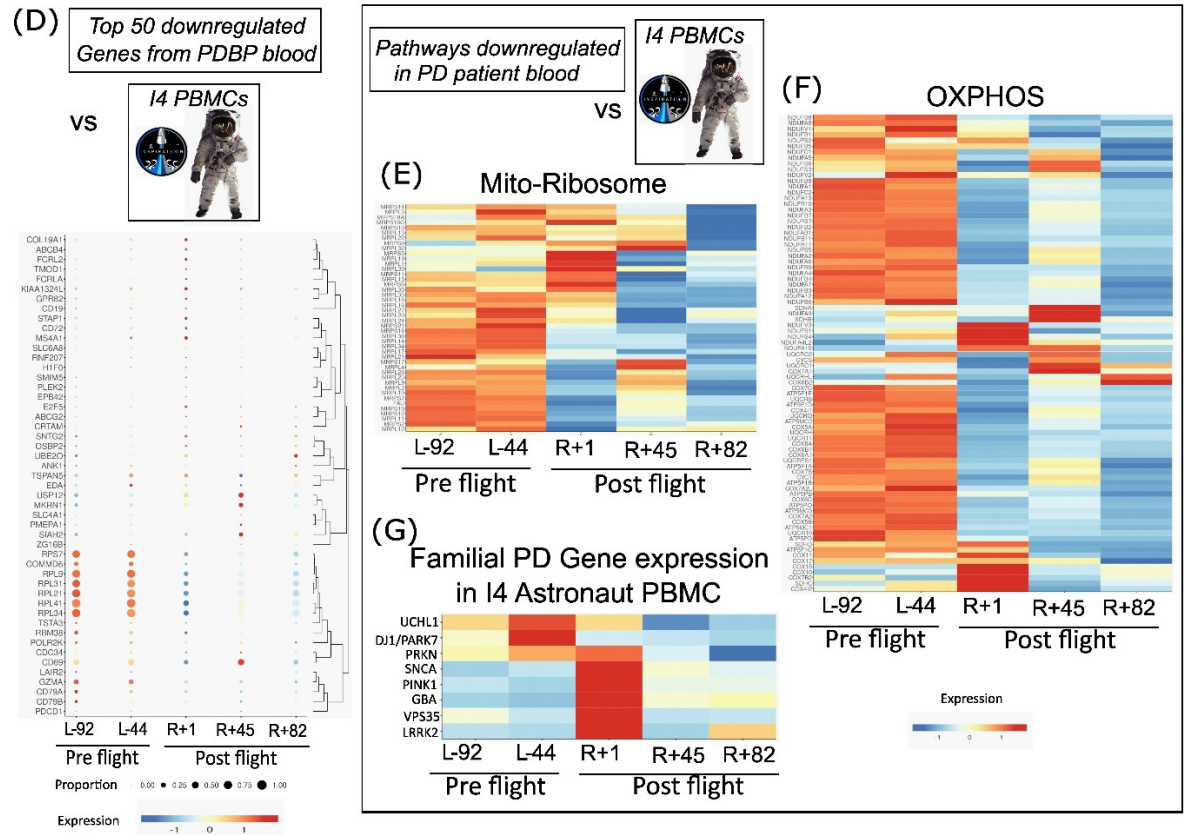

***Fig 1. PD-associated Genes and Spaceflight:*** *Network of genes from Gene Set Enrichment Analyses (GSEA) of human microarray datasets comparing flight to ground conditions.* ***(A)*** *Data sets reanalyzed from GeneLab (OSDs-13, 52, 114, 174) used for GSEA analysis by da Silviera W.A. et al [17]* ***(B)*** *142 differentially expressed genes (DEGs) from the Parkinson's Disease node of Oxidative phosphorylation cluster [17] (derived from Fig 2A from da Silveira WA et al). Nodes are individual DEGs with outer color representing participation in respective diseases (KEGG pathways), with FDR values and number of genes. Edges shown in confidence view with active interaction sources from experiments, databases, co-expression, neighborhood, gene fusion and co-occurrence[29] (interaction score > 0.4).* ***(C)*** *Pie chart showing the molecular functions (left) and Biological processes (right) of the genes, presented as % terms per group.* ***(D)*** *Bubble plots for 25 out of top 50 downregulated genes in PD identified from PDBP blood data [21] as observed in Inspiration 4 (I4) astronaut blood PBMCs.* ***(E)*** *Expression heatmaps for pathways significantly downregulated in PD such as genes from mitochondrial ribosome and* ***(F)*** *OXPHOS were also downregulated in I4 astronaut's blood PBMCs post flight.* ***(G)*** *Familial PD gene expression were also found to be altered in I4 astronaut's blood post flight. All I4 data were generated from Inspiration4 Multiome Data Explorer, PBMCs [22, 23]. Flight timeline denoted as L= Launch or R=Return followed by the number of days.*

**Changes in metabolites**

*Homocysteine*

Recent findings indicate that spaceflight and PD also share visual symptoms, specifically 'Spaceflight Associated Neuro-ocular Syndrome' (SANS), which is akin to the visual motion hypersensitivity (VMH) reported by some PD patients [30]. The underlying mechanism for VMH and SANS is linked to MD and dopamine (DA) depletion [30, 31]. Zwart et al. showed that astronauts affected with SANS had higher plasma homocysteine and iron levels post spaceflight [32]. Similarly, plasma homocysteine in PD patients is elevated compared to that of healthy individuals, and this high homocysteine level is involved in MD, neuronal apoptosis, oxidative stress, and DNA damage [33]. Additionally, elevated integrated stress response signaling is observed in the brains of patients and animal models of PD [34] as well as post-spaceflight [17].

*Dopamine*

PD is characterized by a loss of dopaminergic neurons in SNpc, which results in a decrease in striatal DA. Data from post-flight astronauts suggest decreased levels of DA metabolites, homovanillic acid (HVA) and 3-Methoxytyramine (3-MT), in urine and cerebrospinal fluid (CSF), indicating lower DA levels. Simulated microgravity experiments using *C. elegans* proved that the loss of mechanical contact stimuli in microgravity elicits decreased DA and *comt-4* (catechol-O-methyl transferase) expression; and the animals displayed reduced movement [35]. These results corroborate findings by Popova et al., who showed that space-flown mice had significantly decreased expression of a key DA biosynthesis gene (TH), and genes involved in DA metabolism (MAOA) and O-methylation (COMT), in the SNpc region of the brain [5, 36], adding to the list of Parkinsonism features post-spaceflight.

*Vitamin D*

Deficiency in serum vitamin D (25(OH)-D) is prevalent among PD patients, with levels corresponding to the risk of falls and non-motor symptoms [37]. Astronauts, too, experience significant decreases in both 25(OH)-D and 1,25(OH)$_2$-D forms of vitamin D, for which they take vitamin D3 supplements during missions [17]. It is important to note that Vitamin D3 is modified by liver mitochondria into its active form, 1,25 vitamin D, and systemic MD may contribute to the reduced levels [17].

**Physiological parallels**
Both PD and spaceflight include complaints of olfactory dysfunction (OD) and sleep disturbances. In PD, OD serves as a hallmark for early-stage diagnosis [38], while astronauts experience altered senses of taste and smell during (and post) spaceflight [17]. PD patients present symptoms of REM sleep disorder [37, 38], and astronauts report space-insomnia including shortened sleep duration, reduced sleep quality, and an increased difficulty in falling or staying asleep [17]. Additionally, both spaceflight and PD present with similar changes in the gut microbiome, with overall dysbiosis and a decrease in short chain fatty acid synthesizing bacteria [39].

**Systemic inflammation**
Neutrophil to Lymphocyte ratio (NLR) is regarded as a hallmark biomarker for systemic inflammation and stress, and there is a significantly increased NLR, due to increase in neutrophil count and lower lymphocyte counts in PD patients, throughout the disease course [40]. Spaceflight leads to an increase in NLR, however, it attains normalcy within a week after landing [41].

**Similarities with other neurodegenerative disorders**
While the increased risk to astronauts of Alzheimer's disease (AD) and dementia are being studied [42, 43], to our knowledge this is the first evidence-based study to link PD and long-term spaceflight. Jaster et al. opined that spaceflight results in brain ischemia and a higher occurrence of PD than in the general population [44, 45], however they did not provide convincing epidemiological or clinical data for this opinion. Because there are molecular similarities between PD and other neurodegenerative disorders such as AD and HD **(Table 1)** more research is needed to establish a specific link to PD. This is challenging because PD and related diseases, collectively referred to as Parkinsonism, share common cardinal movement symptoms but are heterogeneous in their molecular signatures, such as presence or absence of Synucleinopathy [1]. Although the presence of alpha synuclein inclusions post space-flight has not been investigated, it has been reported that 6 months after exposure to Galactic Cosmic Radiation (GCR) there was an increase in Aβ plaque accumulation and cognitive impairment in an AD mouse model [43].

We hypothesize **(Fig 2)** that the stressors of spaceflight consistently lead to systemic mitochondrial dysfunction [17] and elevated inflammation [41] resulting in structural changes in the BG [6, 8], loss of TH in the SN, and decreased DA neurotransmitter [5, 13, 36]—which leads to altered gait [12]. However, other regions of the brain exhibit greater plasticity and may return to normal function several months after spaceflight [7]. Nevertheless, the sustained changes in the BG, which includes the SNpc and Str, pose a risk to astronauts for developing movement disorders during and post prolonged deep space travel. These brain regions house the most vulnerable neurons, the degeneration of which underlies at least two neurodegenerative movement disorders: PD [20] and HD [46].

**Conclusion**
The list of analogous features **(Table 1)** are not intended as a diagnostic, but shows that the concerns raised in this paper extend to other neurodegenerative disorders with overlapping symptoms. The parallels between spaceflight and Parkinsonism raise significant concerns regarding the potential development of Parkinsonism-like conditions during prolonged deep space missions. While some shared characteristics may resolve over time, the risk of severe neurological damage in deep space is potentially catastrophic. As humanity sets its sights beyond low-Earth orbit to the Moon and Mars missions [17], it is imperative to understand and mitigate these neurological hazards.

Given the absence of a cure for PD, preventive measures are essential to safeguard space explorers. Proactive antioxidant therapy [41] and mitochondrial protections are sensible, considering the link between MD and PD. Studies focusing on mitochondria-targeted therapies (pre and during spaceflight) would help mitigate these concerns [17, 31]. Furthermore, emerging evidence suggests that α-synuclein seed amplification assays (SAAs) may enable the early diagnosis of PD, years before the onset of classical symptoms and significant neuronal loss in the SNpc [38]. Recently a diagnostic tool has been reported for detection of phosphorylated α-synuclein from skin biopsies of individuals affected with Synucleinopathy such as PD, Multiple System atrophy, Lewy body dementia and pure autonomic failure [47]. Monitoring these changes could allow astronauts to assess the risk of PD and related disorders.

PD represents a potential long-term health concern for astronauts on deep space missions. Understanding the interplay between spaceflight stressors, MD, and PD could help mitigate damage, and benefit those on Earth who are affected by, or at risk for, PD.

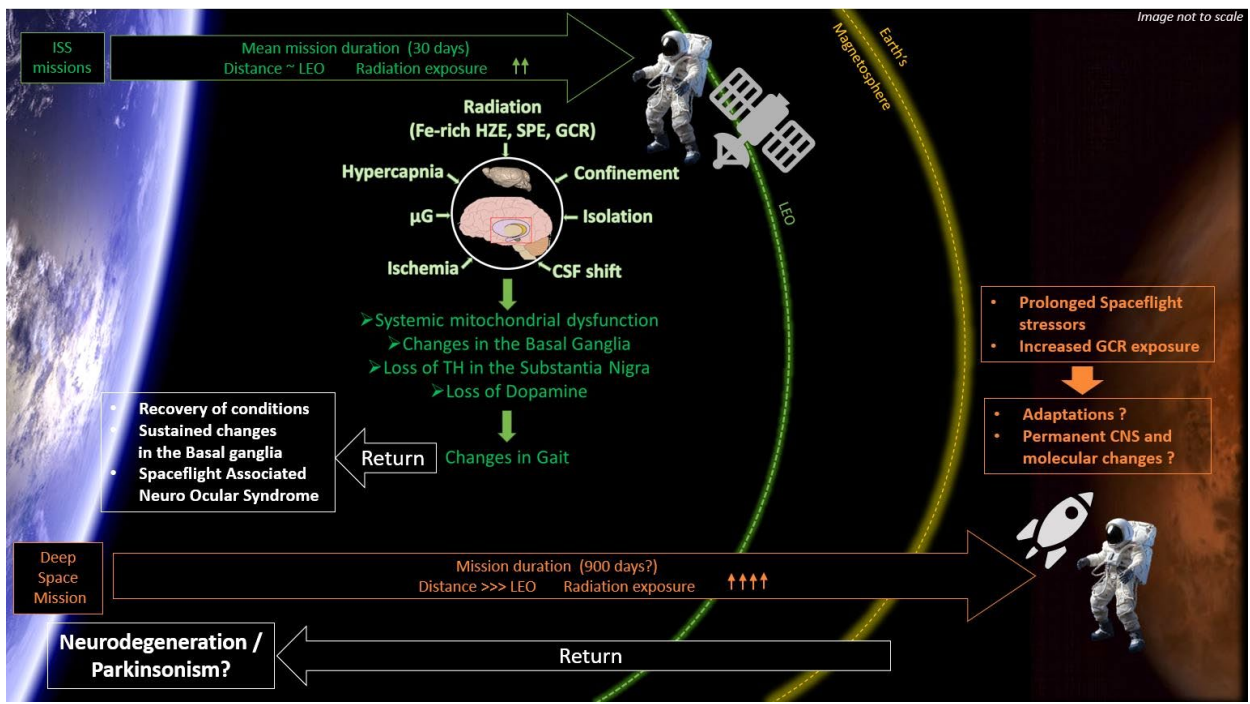

*Fig 2:* Schematic representation showing the effect of indicated spaceflight stressors revealed from studies on human and mouse brain from missions within the low earth orbit (LEO) International Space station (ISS). Post spaceflight there is a sustained increased in the Basal Ganglia (BG) volume several months post-spaceflight, loss of Tyrosine hydroxylase (TH) gene expression and Dopamine neurotransmitter levels in the Substantia Nigra pars compacta and Striatum, and systemic mitochondrial dysfunction with

*Parkinson's disease (PD) pathway genes affected. These molecular and physiological changes in the Central Nervous System (CNS) may lead to the observed PD-like changes in gait and muscle loss, which fade over time. However, there are some sustained changes such as Spaceflight Associated Neuro Ocular Syndrome and changes in the volume of BG. For deep space missions beyond LEO and Earth's magnetosphere, the radiation exposure and mission duration will be drastically increased. This could lead to permanent damage with neurodegenerative disorders such as Parkinsonism.*

**Table 1**
*Molecular and physiological parallels observed between spaceflight and neurogenerative diseases such as PD, HD and AD.*

| | Factors | Spaceflight (SF) | Parkinson's disease (PD) | Huntington's disease (HD) | Alzheimer's disease (AD) |
|---|---|---|---|---|---|
| 1 | Etiology | Spaceflight stressors such as microgravity, cosmic radiation, hypercapnia, ischemia, isolation, confinement etc. Genetics may influence outcome [8, 16, 17] | Genetic due to mutations in SNCA, LRRK2, PARK2, GBA and other genes. Environmental components including toxin exposure, lifestyle, stress, and aging contributes as well [1, 3] | Genetic with autosomal dominant inheritance in HTT gene due to Trinucleotide Repeat (CAG) Expansion [46] | Genetic due to mutation in APP, PSEN1, PSEN2 and APOEe4 gene. Environmental factor, lifestyle, TBI, chronic inflammation and aging as contributing factors [48] |
| 2 | Movement changes | Changes in gait, muscle loss, reduced muscle strength in and post-flight; recovered with time [12, 17, 35, 49] | Changes in muscle rigidity, gait, resting tremor, slowness in movement [20] | Presence of chorea, rigidity, dystonia, dysphagia, and akinesia [46] | Diminished movement coordination [48] |
| 3 | Mitochondrial dysfunction (MD) | Systemic MD observed post SF such as dysregulation of OXPHOS and ETC complexes [16-18] | MD is a key hallmark of PD in brain (SNpc), blood, and muscles [20] with dysfunction in OXPHOS [20, 26], ETC complexes (I and IV) [25, 26], mito-biogenesis, and mitophagy; suspected to be causative [24] | Present in striatal medium spiny neurons; ETC Complex II and IV [46] | MD present as effect of aging and tau oligomer-induced damage [48] |
| 4 | Proteostasis | Failure in proteostasis observed post SF [17] | Loss of proteostasis function observed in PD brain and animal models [27, 28] | Disruption of proteostasis present [46] | Failure in Proteostasis and Ribostasis reported [28, 48] |
| 5 | Protein aggregation | GCR increased Amyloid-beta (Aβ) aggregation in AD mice model [42] | α Synuclein misfolds and aggregates to form intracellular inclusions or Lewy bodies [1, 26, 28] | HTT protein with expanded poly glutamine (CAG) repeats misfolds and aggregate in neurons [46] | Aβ peptides aggregate to form extracellular plaques. Hyperphosphorylated Tau forms neurofibrillary tangles inside neurons [48] |
| 6 | The integrated stress response (ISR) | Elevated ISR is anticipated post SF [17] | Elevated ISR signaling in the brain of PD and AD patients; and in animal models of neurodegeneration, prion disease, traumatic brain injury, and myelination disorders [34, 46, 48] | | |

| # | | | | | |
|---|---|---|---|---|---|
| 7 | Serum Homocysteine level | Elevated [31, 32] | Elevated [33] | Limited evidence | Elevated [48] |
| 8 | Vitamin D levels | Decreased [17] | Decreased [37] | May decrease | Decreased [48] |
| 9 | Changes in neurotransmitters (NTs) | DA & its metabolite HVA, 3-MT levels decreased [5, 35, 36] | DA & its metabolite HVA, 3-MT levels decreased as hallmark of PD [20, 25] | Decreased GABA [46]. DA & its metabolites altered with early increases followed by late decreases [50] | Acetylcholine altered primarily [51]; Alterations to DA reported through loss of VTA neurons [52]. |
| 10 | Changes in brain | Presence of PVS burden; sustained change in BG and Ventricles [6, 7] and reversible changes in cortex [10, 11] | Change in BG circuitry, death of Nigrostriatal projections and SNpc neurons; Presence of PVS burden; [4] | Change in BG circuitry, Str; medium spiny neurons and cortical projections affected [46, 50] | Hippocampus, amygdala, cerebral cortex, VTA [48, 52] |
| 11 | Neutrophil to Lymphocyte Ratio | Increased during SF [41] | Increased in PD patients throughout disease course [40] | Limited evidence | Reported as a predictor of AD and related dementia [53] |
| 12 | Vision changes | SANS associated post SF [31] | VMH associated with PD pathology [30] | Present [46] | Presence of ocular motility dysfunctions and poor visual attention [51] |
| 13 | Olfactory changes | Alterations in smell and taste during SF [17] | Olfactory dysfunction common in PD, before PD diagnosis [38] | May be present | May be present [48] |
| 14 | Changes in sleep | Space insomnia, circadian desynchronization, common during and post SF [17] | Rapid eye movement (REM) sleep behavior disorder (RBD) or hyposmia common in PD [37, 38] | Lower sleep efficiency, shortened REM sleep [46] | Insomnia, sleep-disordered breathing, sleep fragmentation and sleep-related movement disorders associated with a higher risk of AD [54] |

A list of abbreviation is listed below:

1. 3 MT - 3 Methoxy Tyramine
2. Aβ - Amyloid-beta
3. AD - Alzheimer's disease
4. APP - Amyloid Precursor Protein
5. APOEe4 - Apolipoprotein E epsilon 4 allele
6. a-Syn – Alpha synuclein
7. BG - Basal Ganglia
8. COMT - Catechol-O-Methyltransferase
9. COMT 4 - Catechol-O-Methyltransferase 4
10. CSF - Cerebrospinal Fluid
11. DA - Dopamine
12. DEG - Differentially Expressed Genes
13. ETC - Electron Transport Chain
14. GABA - Gamma Amino Butyric Acid
15. GBA - Glucocerebrosidase
16. GCR – Galactic Cosmic Radiation
17. GSEA - Gene Set Enrichment Analysis
18. HD - Huntington's disease
19. HTT - Huntingtin
20. HVA - Homovanillic Acid
21. ISR - Integrated Stress Response
22. LEO - Low-Earth Orbit
23. LRRK2 - Leucine-rich repeat kinase 2
24. MAO A - Monoamine Oxidase A
25. MD - Mitochondrial Dysfunction
26. MPS – Million Person Study
27. NLR – Neutrophil to Lymphocyte Ratio
28. NT - Neurotransmitter
29. OD - Olfactory Dysfunction
30. OXPHOS - Oxidative Phosphorylation
31. PARK2 - Parkin RBR E3 ubiquitin protein ligase
32. PD - Parkinson's Disease
33. PSEN1 - Presenilin 1
34. PSEN2 - Presenilin 2
35. PVS - Perivascular Space
36. RBD - Rapid Eye Movement (REM) Sleep Behavior Disorder
37. REM - Rapid Eye Movement
38. RR – Rodent Research
39. SAA - Seed Amplification Assay
40. SANS - Spaceflight Associated Neuro Ocular Syndrome
41. SNCA - Synuclein Alpha
42. SNpc - Substantia Nigra Pars Compacta
43. Str - Striatum
44. TH - Tyrosine Hydroxylase

45. VMH - Visual Motion Hypersensitivity
46. VTA - Ventral Tegmentum Area